\documentclass[envcountsame]{mmnp}
\usepackage[tbtags]{amsmath}
\usepackage{amssymb}
\usepackage{graphicx}
\usepackage[latin1]{inputenc}

\idline{Vol. 8, No. 2}{144}
\doi{10.1051/mmnp/20138210}

\numberwithin{equation}{section}

\begin{document}

\title{Rocking
Subdiffusive Ratchets: Origin, 
Optimization and Efficiency}


\author{I. Goychuk \inst{1} \thanks{\email {goychuk@physik.uni-augsburg.de}} 
\sep V. O. Kharchenko \inst{2} 
 }

\vspace{0.5cm}

\institute{\inst{1} Institute of Physics, University of Augsburg, 
Universit\"{a}tstr. 1, D-86135 Augsburg, Germany \& \\ 
Institute for Physics and Astronomy, University of Potsdam, 
Karl-Liebknecht-Str. 24/25,\\
D-14476 Potsdam-Golm, Germany \\
\inst{2} Institute of Applied Physics, NAS of Ukraine, 58 Petropavlovskaya str., 40030 Sumy,
 Ukraine }


\abstract{We study origin, parameter optimization, and thermodynamic efficiency of isothermal rocking 
ratchets based on fractional subdiffusion 
within a generalized non-Markovian Langevin equation approach. A corresponding 
multi-dimensional Markovian embedding dynamics is realized using a set of auxiliary Brownian particles
elastically coupled to the central Brownian particle (see video on the journal
web site).  We show that anomalous subdiffusive transport
emerges due to an interplay of nonlinear response and viscoelastic effects for fractional
Brownian motion in periodic potentials with broken space-inversion symmetry and driven
by a time-periodic field. 
The anomalous transport becomes optimal for a subthreshold driving 
when the driving period matches a characteristic time scale of interwell 
transitions. It can also be optimized by varying temperature, amplitude of periodic potential
and driving strength. The useful work done against a load shows a parabolic dependence on the
load strength. It grows sublinearly with time and
the corresponding thermodynamic efficiency decays 
algebraically in time because the energy supplied by the driving field scales with time linearly.
However, it compares well with the efficiency of normal diffusion rocking ratchets
on an appreciably long time scale.  }





\titlerunning{Rocking subdiffusive ratchets}

\maketitle


\vspace*{1cm}

\setcounter{equation}{0}
\section{Introduction}

Theories of anomalous Brownian motion stem from pioneering
contributions by Kolmogorov \cite{Kolmogorov}, Mandelbrot and van Ness 
\cite{Mandelbrot}, Montroll and Weiss \cite{Montroll}, 
Shlesinger \cite{Shlesinger1}, Scher and Montroll \cite{Scher},
and others. Mandelbrot and van Ness coined first the notion
of fractional Brownian motion (fBm) featured by long range interdependence of the particle position 
increments, although fBm was implicitly introduced earlier by Kolmogorov as a class of curves
in an abstract  Hilbert space. They 
considered a Riemann-Liouville fractional integral of the white Gaussian noise $\xi(t)=
dB(t)/dt$, or the Holmgren-Riemann-Liouville fractional integral of normal Brownian motion $B(t)$,
\begin{equation}
 \label{1}
 B_H(t,t_0)=\frac{\tau_0^{-H+1/2}}{\Gamma(H+1/2)}\int_{t_0}^t(t-t')^{H-1/2}\xi(t')dt',
 \end{equation}
with $t_0\to -\infty$ to ensure the stationarity of increments. The parameter $H$ is the Hurst exponent 
$0<H<1$ and 
an arbitrary time scaling factor $\tau_0$ guarantees that the physical dimension of $B_H$, e.g. that of
coordinate $x=B_H$, is the same for any $H$.
Such an integral diverges, but the difference $B_H(t):=B_H(t,-\infty)-B_H(0,-\infty)$
remains finite. It serves as a definition for fBm. The variance of fBm scales in time as $Var(B_H(t))
\propto t^\alpha$. Here, $\alpha =2H$ is the power exponent of anomalous diffusion. The self-similar
scaling of normal Brownian motion, i.e. the fact that the processes $h^{-1/2}B(ht)$ and 
$B(t)$ are identical in distribution implies akin equivalence between the processes $h^{-H}B(ht)$ and
$B_H(t)$, in accordance with the above definition. The case of $H=1/2$ ($\alpha=1$) corresponds
to normal diffusion with independent position increments. Slowly decaying negative correlations
for $0<H<1/2$ lead to subdiffusion, $0<\alpha<1$, while the positive ones for $1/2<H<1$ yield
superdiffusion, $1<\alpha<2$.

 Fractional Brownian motion can emerge \cite{GH07,Goychuk09,Goychuk12a} 
as a solution of the
generalized Langevin equation or GLE \cite{Kubo1,Kubo2,Zwanzig1,Zwanzig2}
\begin{equation}
 \label{2}
 m\ddot x + \int_{-\infty}^t \eta(t-t')\dot x(t')dt'=f(x,t)+\xi(t),
 \end{equation}
where the memory kernel $\eta(t)$ and colored zero-mean Gaussian noise $\xi(t)$ obey the 
fluctuation-dissipation
relation,
\begin{equation}
 \label{3}
 \langle\xi(t)\xi(t')\rangle=k_BT\eta(|t-t'|)\;
 \end{equation}
at the temperature $T$ which characterizes the thermal environment and the intensity of the thermal 
noise it produces. 
Brackets denote averaging over the noise realizations, and $k_B$ is the Boltzmann constant. 
The following conditions are required for this: (1)
The corresponding noise is fractional Gaussian noise (fGn) defined as the derivative of
fBm, $\xi(t)=\dot B_{1-\alpha/2}(t)$. \footnote{ Notable is that the superdiffusive persistent noise source
in GLE (\ref{2}) yields anti-persistent subdiffusion as the GLE solution and vice versa. 
This is due to the influence of memory kernel, integral of which diverges
for subdiffusion and is zero for superdiffusion, in the limit $t\to\infty$. }
(2) The memory kernel is accordingly
\begin{equation}
 \label{4}
 \eta(t)=\frac{\eta_{\alpha}}{\Gamma(1-\alpha)}t^{-\alpha}
 \end{equation}
for $0<\alpha<1$, where $\Gamma(z)$ is the gamma-function. 
For $1<\alpha<2$ and $t>0$, one defines the memory kernel by the same expression, which takes on
negative values. However, there is a positive singularity at $t\to 0$, so that the 
integral of the memory friction is always positive for such a superdiffusive GLE and it tends to zero
with the upper integration limit $t\to \infty$. (3) The inertial effects are
negligible, $m\to 0 $ ($m$ is the mass of Brownian particle), and (4)  an external regular force is
absent, 
$f(x,t)\to 0$. This particular 
model is also known under the label of fractional Langevin equation (FLE) 
\cite{Mainardi96,Lutz,GH07}
upon the formal use of the operators of fractional derivative to abbreviate the frictional memory term.
The physical origin  of subdiffusion for $0<\alpha<1$ is viscoelasticity \cite{Goychuk09,Goychuk12a}. 
The memoryless
kernel $\eta(t)=2\eta_1\delta(t)$, where $\eta_1$ is the viscous friction coefficient, yields normal
Brownian motion. It corresponds to a viscous Newtonian fluid, which can be characterized by a macroscopic
viscosity coefficient $\zeta$, $\eta_1\propto \zeta$. Another extreme of non-decaying memory kernel, 
$\eta(t)\approx const$, corresponds to a quasi-elastic trapping force. Maxwell was  first 
\cite{Maxwell} to discuss within a purely macroscopic setting, which  neglects fluctuations, the
emergence of the phenomenon of viscosity of fluids from the elasticity of solids by letting the elastic
deformation forces relax exponentially in time. This corresponds to a simple viscoelastic fluid, which
can be characterized by a viscosity kernel exponentially decaying in time. The corresponding generalized
Brownian motion is characterized by the memory kernel $\eta(t)=\kappa \exp(-\nu t)$, where $\kappa$ 
is an elastic constant and $\nu$ is a corresponding relaxation rate. 
Gemant found \cite{Gemant} that  power law dependences
like one in (\ref{3}) describe well a number of complex viscoelastic materials. Benchmark of a complex
fluid is, however, that its macroscopic viscosity is finite. This implies that a total 
integral of the memory
kernel $\tilde \eta(0)$, where $\tilde \eta(s)=\int_0^t \exp(-s t)\eta(t)dt$ is the Laplace-transformed 
memory kernel, is finite as well. Therefore, a long range memory cutoff $t_{\rm h}$ is necessarily present for any 
liquid-like matter.
The existence of an upper bound $\omega_{\rm h}$ for the frequency of vibrational modes of any medium 
implies also the existence of a short memory cutoff $t_{\rm l}\propto 1/\omega_{\rm h}$. 
For this reason, a power law scaling leading to subdiffusion
always extends in nature only over some finite number, 
 $N_{\rm max}=\log_{10}(t_{\rm h}/t_{\rm l})$, of time, or frequency
decades, and anomalous Brownian motion is mostly a transient phenomenon. However, it can be  a dominant
mechanism of transport and diffusion on mesoscale \cite{Goychuk12b}
which makes the mathematical models of fBm and FLE
 very important.
 Any fractal
scaling in nature is restricted by a certain maximal spatial and/or time range, mostly 
over several 
decades only, seldomly over six or more spatial, or time decades. 
This fact does not undermine, however, grandeur and usefulness of the 
fractal type modeling approaches \cite{MandelbrotBook}.

Montroll and Weiss \cite{Montroll}, Shlesinger \cite{Shlesinger1}, and Scher 
\cite{Scher} proposed a different
description of anomalous Brownian motion based on continuous time random walks (CTRWs) with
a heavy-tail distribution of the residence times spent on the sites of localization before
making a spatial jump to a different localized state. Such a CTRW is 
semi-Markovian, i.e. the residence times on different sites are independently distributed
and the spatial steps are not correlated,
but the first moment of the residence time distribution (RTD)
can diverge, as e.g. for $\psi(\tau)\propto \tau^{-1-\alpha}$, $0<\alpha<1$. This divergence gives
rise to subdiffusion, if the spatial distribution of the particle's jumps has first two moments 
finite. Such a subdiffusion mechanism can be related to a spatial and/or energy disorder. Namely,
RTDs are exponential and can be characterized by a spatially dependent transition rate. However, the 
escape rates out of different localized traps are  vastly different due to disorder, 
so that the averaging of the RTD over disorder yields an effective RTD with divergent mean
time. In the effective medium or mean-field approximation, the later approach is equivalent 
to a semi-Markovian CTRW \cite{Bouchaud,Hughes}. Then, a
spatial disorder is equivalent to a stochastic clock with vastly 
varying but independent time intervals elapsed between jump events (as usually, mean field
approaches neglect correlations). In the continuous space limit, this 
another anomalous Brownian motion can be described by the fractional diffusion equation and its
extension to the diffusion in external force-fields, the
fractional Fokker-Planck equation \cite{Metzler,Barkai01}. It
is related to normal Brownian motion in accordance with the concept of subordination 
\cite{Barkai01,Sokolov}. One can obtain it from the normal one by replacing the physical
time $t$ with a stochastic intrinsic time $\tau(t)$, $B(t)\to B(\tau(t))$. 
$ B(\tau(t))$ is thus a doubly stochastic process. Intrinsic stochastic time, or the
particle's ``age`` is measured by a number of 
random clock periods accomplished within the time interval 
$[0,t)$. Notice that the occurrence of very long time periods implies smaller $\tau$. 
For this reason the average of $\tau$ exists and it scales with physical time as
$\langle \tau\rangle \propto t^\alpha$. This random time is described by a probability density
$p_\alpha(\tau|t)=\frac{t}{\alpha \tau^{1+1/\alpha}}{\cal L}_{\alpha}(t/\tau^{1/\alpha})$, where 
${\cal L}_{\alpha}(x)$ is the probability density of one-sided Levy stable distribution 
\cite{Barkai01,He,Sokolov09,GH11}.
For normal diffusion, $p_1(\tau|t)=\delta(t-\tau)$, i.e. $\tau$ is not a random
variable. It has the fixed value $t$. In an infinite medium,
the averaging over one particle trajectory
yields a normal diffusion scaling of the position variance with time.
It can be characterized by a normal diffusion coefficient which, however, depends on the time interval
of observation and algebraically decays upon the observation time increase \cite{Lubelski,He}. 
Different trajectories
are characterized by different normal diffusion coefficients, but the ensemble averaging
yields subdiffusion. This indicates clearly a breaking of ergodicity and is very different from
the case of fBM subdiffusion, where a single-particle trajectory averaging yields subdiffusive behavior
for each particle separately, and which  is clearly ergodic \cite{Deng,Goychuk09}.
Averaging of such a doubly stochastic process over the realizations of $\tau$
yields a random process
\begin{equation}
 \langle B_{\alpha, \rm CTRW}(t)\rangle_\tau=\int_{0}^\infty p_\alpha(\tau|t)B(\tau) d\tau \;.
 \end{equation}
It is very different from fBm, even though the scaling of its variance with time is virtually the same.

In the focus of this work is the noise-assisted anomalous subdiffusive ratchet transport \cite{G10} in a 
spatially periodic force-field
with broken space inversion symmetry and driven in addition by a periodic external force
\cite{Magnasco,Bartussek,Reimann,Marchesoni}. 
At thermal equilibrium,
the net transport is absent because the symmetry of thermal detailed balance
forbids such a transport if only the transport is not induced by a traveling potential trapping the
particle and carrying it on. The latter transport has almost mechanical character and is not induced 
but rather hindered
by thermal fluctuations. It is not considered here. 
Furthermore, the noise-assisted ratchet transport is absent in the linear response regime
to periodic driving due to the Onsager symmetry
of kinetic coefficients for a nearly equilibrium transport. A net rectification current  
can emerge only in a strongly nonlinear and non-equilibrium regime of driven transport. This
general circle of problems is known under the label of  Brownian ratchets 
physics \cite{Reimann,Marchesoni}. 
A huge number of papers is
devoted to normal diffusion ratchets as reviewed in \cite{Reimann,Marchesoni}. 
However, a generalization  to
the domain of anomalous diffusion is highly nontrivial.  Physics of such anomalous
Brownian motors is currently still in its infancy.
The problem is that within a CTRW subdiffusion scenario based on the RTDs with divergent mean 
trapping time the asymptotic
response to a periodic driving is absent \cite{Barbi,SokolovKlafter06,H07,H09}, 
contrary to some published papers
which implicitly assumed that the external field oscillates in the intrinsic random time $\tau$ and
not the laboratory time $t$. 
This death of response not only in the linear response regime 
\cite{SokolovKlafter06}
but also beyond \cite{H07,H09} makes the fluctuating tilt or rocking ratchets based
on the CTRW subdiffusion barely possible. However, the viscoelastic subdiffusion of the fBm type
also displays a remarkable feature which makes the emergence of rectification effect highly nontrivial.
Namely, neither diffusion nor transport depend asymptotically on the presence of 
a periodic potential \cite{Goychuk09,GH11,Goychuk12a}.
This makes ratchet effect for an adiabatically slow driving with vanishingly small frequency clearly
impossible. The transient to this asymptotic transport regime is, however, very slow. It strongly
depends, in particular, on the amplitude of the periodic potential and 
temperature \cite{Goychuk09,GH11,Goychuk12a} which makes anomalous ratchet effect possible 
in a nonadiabatic driving regime \cite{G10}.
Moreover, the anomalous transport optimizes when the periodic tilts synchronize 
with an intrinsic time scale of activated transitions to nearest potential wells \cite{GKh12a}. 
An optimal anomalous transport
occurs as a manifestation of non-Markovian stochastic resonance \cite{Gammaitoni,GH03,GH04,GH05}. 
Moreover, it is a genuine ratchet effect
since the transport persists against a sufficiently small loading force and there exists a critical stopping
force which depends on the model parameters \cite{GKh12a}. 

In this paper, we continue to investigate the optimization of anomalous transport
depending on the model parameters such as temperature, potential height, frequency 
and strength of 
the driving field. Moreover, we address for the first time the question of thermodynamic efficiency of anomalous
isothermal Brownian
motors.

\section{Theory}

\begin{figure}[htbp]
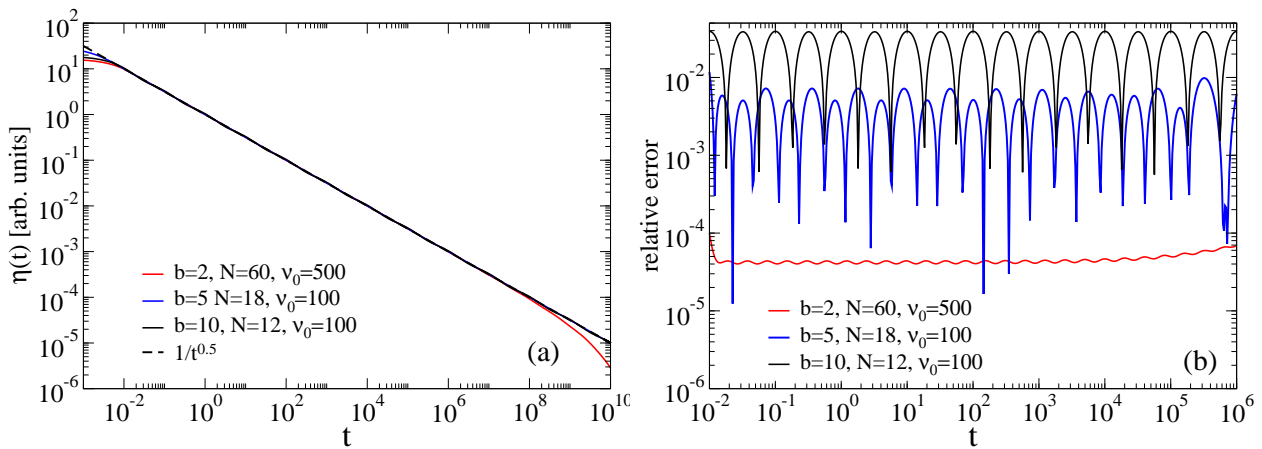

\includegraphics[scale=0.33]{fig1a.eps}
\includegraphics[scale=0.33]{fig1b.eps}
\caption{(a) Power law kernel, $1/t^\alpha$, \textit{vs.} its approximation by a sum of exponentials,
$C_\alpha(b)\sum_{i=1}^N \nu_i^\alpha\exp(-\nu_i t)$, $\nu_i=\nu_0/b^{i-1}$, for  $\alpha=0.5$ and different 
$b$, $\nu_0$, and $N$, with $C_{0.5}(10)=1.3$,  $C_{0.5}(5)=0.909$, $C_{0.5}(2)=0.39105$; 
(b) relative error of the corresponding approximations. }
\label{fig1}
\end{figure}

\begin{figure}[htbp]

\centerline{\includegraphics[scale=0.45]{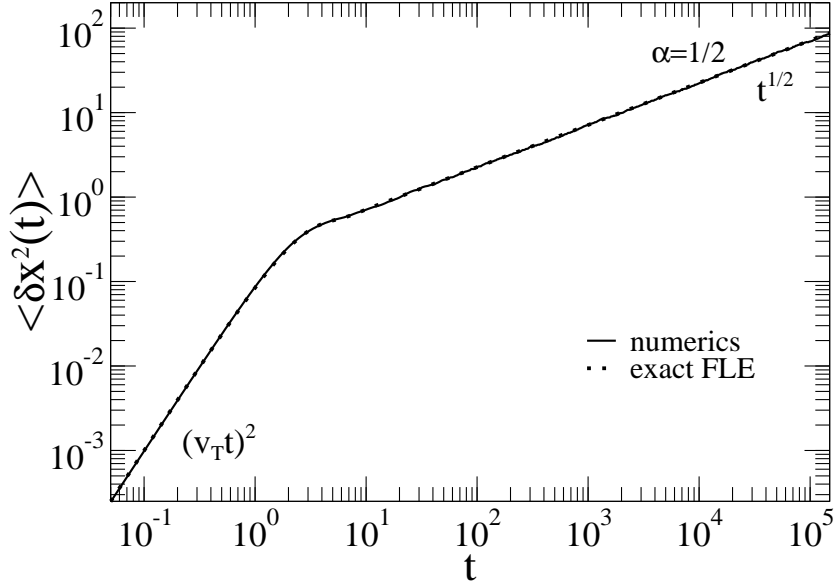}}
\caption{Numerics \textit{vs.} exact result for the FLE dynamics without potential and for $T=0.1$. 
Other parameters are given
in the text. The discrepancy is less than 5\% even for large $t$ in this figure. 
Initially diffusion is always
ballistic, as a manifestation of the inertial effects. It changes into subdiffusion
on the time scale $t\gg 1$ exceeding much characteristic time of initial positive velocity autocorrelations
defined by $\tau_v$. Slowly decaying negative velocity autocorrelations yield subdiffusion.}
\label{fig2}
\end{figure}

A nice feature of fBm and its generalization, which includes inertial effects and action of external forces, 
is that it can be derived  from a purely dynamical model following to the reductionist type  dynamical
approach by Bogolyubov \cite{Bogolyubov}, Ford, Kac, and Masur \cite{Ford}, Zwanzig \cite{Zwanzig1},  
and others. 
Indeed, GLE (\ref{2}) follows from a purely Hamiltonian
model of a transport particle 
moving in a time-dependent potential $V(x,t)$ and 
coupled elastically with spring
constants $k_i$ to the harmonic oscillators with masses $m_i$, $H=p^2/(2m)+V(x,t)+\sum_i
[p^2_i/(2m_i)+k_i(x-x_i)^2/2]$. This set of oscillators models the influence of
environment, or thermal bath, see in \cite{Zwanzig2,WeissBook,Kupferman,Goychuk12a} for details. 
The only non-dynamical element in this derivation is
that the initial values of environmental degrees of freedom $\{p_i(0),x_i(0)\}$ are 
random and canonically distributed. 
It is, however,
 quite in the
spirit of modern molecular dynamics. The derivation requires a quasi-continuum  of oscillators
with dense spectrum to ensure that the noise correlation function asymptotically decays 
in time to zero. 
Such  a hyper-dimensional embedding dynamics leading to a
low-dimensional but non-Markovian GLE dynamics is obviously Markovian. Projection
of a multidimensional Markovian dynamics onto a subspace yields typically non-Markovian dynamics of reduced
dimensionality. This makes the whole idea of a 
finite-dimensional stochastic
Markovian embedding of the GLE dynamics obvious. Indeed, let us following to \cite{Goychuk12a} 
consider dynamics of the central particle coupled to a finite set of harmonic Brownian oscillators
modeling the environment.
The corresponding set of Langevin equations reads:
\begin{eqnarray}\label{embedding1}
m\ddot x & =& f(x,t)-
\sum_{i=1}^{N}k_i(x-x_i) \;,\nonumber \\
m_i\ddot x_i& = &k_i (x-x_i)-\eta_i\dot x_i+\sqrt{2\eta_ik_BT}\xi_i(t) \;,
 \end{eqnarray}
where $\xi_i(t)$ are independent unbiased white Gaussian noise sources of unit intensity, 
$\langle \xi_i(t)\xi_j(t')\rangle=\delta_{ij}\delta(t-t')$, and 
$f(x,t)=-\partial V(x,t)/\partial x$. Unlike in the purely dynamical model,  we (i) 
took only a finite set of effective medium's
oscillators, and (ii) replaced the rest by the Stokes frictional forces with frictional
coefficients $\eta_i$, and the corresponding thermal noises acting on these
representative Brownian quasi-particles. Furthermore, let us (iii) neglect the inertial effects for 
the medium's degrees of freedom, $m_i\to 0$, and introduce the (visco)elastic
forces $u_i=-k_i(x-x_i)$ acting on the central particle. Then, the stochastic dynamics 
in (\ref{embedding1}) reduces to \cite{Goychuk09}
\begin{eqnarray}\label{embedding2}
\dot x&=& v\;,\nonumber \\
m\dot v & =& f(x,t)+
\sum_{i=1}^{N}u_i(t) \;,\nonumber \\
\dot u_i& = &-k_i v-\nu_iu_i+\sqrt{2\nu_i k_i k_BT}\xi_i(t) \;,
 \end{eqnarray}
where $\nu_i=k_i/\eta_i$ are the relaxation rate constants of viscoelastic forces.
The last equation in (\ref{embedding2}) recalls an equation for the relaxation of elastic
force introduced by Maxwell \cite{Maxwell} in his pioneering discussion of macroscopic viscoelasticity.
It is augmented by the corresponding Langevin force accounting for thermal fluctuations,
in accordance with the fluctuation-dissipation theorem \cite{Kubo1}. 
Therefore, this model can be named generalized Maxwell-Langevin model of viscoelastic stochastic
dynamics. It is easy to show that the exclusion of the dynamics of $u_i(t)$ variables assuming
that the initial $u_i(0)$ are Gaussian distributed, unbiased and with the variance 
$\langle u_i^2(0)\rangle=k_ik_B T$ leads to GLE (\ref{2}), where $\dot x(t)=0$ for $t<0$, 
i.e. the Brownian particle starts to move at $t_0=0$, and the memory kernel presents
a sum of exponentials:
\begin{eqnarray}\label{kernel}
\eta(t)=\sum_{i=1}^N k_{i} \exp(-\nu_i t)\;.
 \end{eqnarray}
For such a memory kernel, the Markovian embedding (\ref{embedding2}) of non-Markovian GLE dynamics
is exact. However, even a power-law dependence in (\ref{4}) can be approximated  by 
(\ref{kernel})  between two time cutoffs $t_l$ and $t_h$ with a 
controllable accuracy, cf. Fig. \ref{fig1},a. Indeed, 
it suffices
to take for this $\nu_i=\nu_0/b^{i-1}$ and $k_i\propto \nu_i^\alpha$ \cite{Palmer}. 
Here, $\nu_0\sim 1/t_l$ is a high-frequency
cutoff, and the parameter $b$ controls the accuracy of approximation. With this choice
it is easy to show that the corresponding sum $\eta(t,\nu_0,b,N)$ (which depends formally on $\nu_0$,
$b$, and $N$) 
has the scaling property, $\eta(ht,\nu_0,b,N)=h^{-\alpha}\eta(t,h\nu_0,b,N)$. The ``scale-free'' power-law
dependence obeys the scaling $\eta(ht)=h^{-\alpha}\eta(t)$, which establishes the meaning
of ``scale-free''. The approximation has indeed this property provided that it does not depend on $\nu_0$.
Of course, this can only be valid  for $t$ far away from the cutoff boundary values. However, the same
is valid also for any power-law  fractal dependence with cutoffs. The absolute error of approximation
(which exhibits logarithmic oscillation, see in Fig. \ref{fig1},b must also be negligibly small.
Indeed, for $\alpha=1/2$ and
$b=10$, the maximal error of the kernel approximation is about $4\%$, which already suffices 
for most stochastic simulations, where $5\%$ accuracy is a ``golden standard''. For $b=5$,
the maximal error reduces to $0.7\%$ and for $b=2$ becomes smaller than
$0.01 \%$, cf. Fig. \ref{fig1},b. The nice feature is that the number of exponentials
required to approximate the power law over $r$ time decades scales as $N\sim r/\log_{10} b$.
The weak dependences on $r$ and $b$ ensure an excellent numerical approach to integrate fractional
Langevin dynamics on any time scale of practical interest with a well controlled accuracy.
Instead of integrating stochastic integro-differential GLE equation with long-range
correlated fractional Gaussian 
noise, where unfortunately no well-established mathematical results are known on the convergency
and accuracy of the existing stochastic integration algorithms, we replaced it by the stochastic simulations
of an extended multi-dimensional Markovian dynamics with very reliable numerical methods at use.
This is probably the best numerical approach 
at present, which allows to simulate very long trajectories.  
Practically we used stochastic Heun algorithm for SDEs \cite{Gard} (second order 
stochastic Runge-Kutta method). The approach was tested, see in Fig. \ref{fig2}, against
the exact solution of the FLE for the potential-free subdiffusion \cite{Lutz}, 
\begin{eqnarray}\label{varFLE}
\langle \delta x^2(t) \rangle=2v_T^2t^2 E_{2-\alpha,3}
[-(t/\tau_{v})^{2-\alpha}]\;. 
\end{eqnarray}
Here, $E_{\gamma,\beta}(z)=
\sum_{n=0}^{\infty}z^n/\Gamma(n\gamma+\beta)$ is the generalized Mittag-Leffler
function, $v_T=\sqrt{k_BT/m}$ is thermal velocity, and 
$\tau_v=(m/\eta_{\alpha})^{1/(2-\alpha)}$ is a time constant of the anomalous
relaxation of equilibrium velocity autocorrelation function, 
$\langle v(t)v(t') \rangle=v_T^2 E_{2-\alpha}
[-(|t-t'|/\tau_{v})^{2-\alpha}]$, where  $E_{\gamma}(z):=E_{\gamma,1}(z)$
is the Mittag-Leffler function. In simulations, we scale time in units of $\tau_v$, and 
distance in arbitrary units of $L$ (which becomes the spatial period in the case 
of periodic potentials). 
Then, the velocity scales in units of $L/\tau_v$,
energy in units of $m L^2/\tau_v^2$, force in units of $m L/\tau_v^2$ and temperature
in units of $m L^2/(\tau_v^2 k_B)$. We used Markovian embedding
with $b=10$ and $N=12$, $\nu_0=100$, time step $\Delta t=0.002$, and $n=10^4$ particles
in simulations for doing ensemble averaging. Temperature was variable.  Simulations were
done with double precision on a Fermi NVIDIA graphical processor unit (GPU)  which reduced
the computational time by a factor of about 100 with respect to standard CPU computing.
The agreement is very good in Fig. \ref{fig2}, within the statistical error of simulations,
which is smaller than 5\%.
It should be also mentioned that for very large times $t\gg t_h=b^{N-1}/\nu_0$
diffusion and transport become normal. However, since $t_h$ grows exponentially
with $N$ this truly asymptotic regime can be made numerically not reachable and thus
completely irrelevant, as in our simulations. When we speak about asymptotic regime
`$t\to\infty$',
we mean that time $t$ is large but yet much smaller than $t_h$, i.e.
still within the regime of anomalous diffusion and transport.
 However, the obtained results correspond to the true $t\to\infty$ limit within
the FLE description. \\

\subsection{Anomalous ratchet }

We use the outlined approach to study anomalous fluctuation-induced transport in periodic 
ratchet potentials $U(x)$ lacking space-inversion symmetry, i.e. $U(x+L)=U(x)$, where $L$ 
is spatial period, but
$U(-x)\neq U(x+x_0)$, for any shift $x_0$, and driven by a periodic force with period ${\cal T}$,
$f_{\rm ext}(t+{\cal T})=f_{\rm ext}(t)$. 
In this paper, $f_{\rm ext}(t)=A\cos(\Omega t)$, $\Omega=2\pi/{\cal T}$. \\ \vspace{0.5cm}

Moreover, a constant biasing force $f_0$ (load) can be applied in the direction opposite
to the direction of rectified motion.
The total potential is $V(x)=U(x)-xf_{\rm ext}(t)+xf_0$. We are interested in the subtransport which can
be characterized by the mean subvelocity
\begin{equation}
 v_\alpha=\Gamma(1+\alpha)\lim\limits_{t\to\infty}\frac{\langle x(t)\rangle}{t^\alpha}.
\end{equation}
The subdiffusion coefficient is defined in a similar manner,
\begin{equation}
 D_\alpha=\Gamma(1+\alpha)\lim\limits_{t\to\infty}\frac{\langle \delta x^2(t)\rangle}{2 t^\alpha},
\end{equation}
where $\langle \delta x^2(t)\rangle$ is the position variance, $\langle \delta x^2(t)\rangle=\langle 
x^2(t)\rangle-\langle 
x(t)\rangle^2$.  
It should be mentioned that the normal velocity (with $\alpha=1$) is zero, $v_1=0$. This does
not mean, however, that the particles are localized and the transport is absent.
In the absence of a periodic driving and/or bias, there cannot be any  directed
transport, including anomalous one. It is forbidden by the symmetry of thermal detailed 
balance \cite{Magnasco,Reimann,Marchesoni}. 
Transport cannot emerge also 
within a linear response regime of periodic driving. However, it can emerge already
as a quadratic
response to such a driving \cite{Reimann}. Indeed, this is 
the case also for anomalous transport. Generally, the emergence of such a rectification
current presents a strongly nonlinear and nonequilibrium effect. If driving is subthreshold, then
a particle initially localized in a potential well would remain trapped in this potential well
in the absence of thermal noise, for a nonresonant driving and sufficiently strong friction. 
 Thermal fluctuations help to overcome
the potential barrier and induce directional transport  of particles. In the case of normal diffusion
such a thermal fluctuation-assisted transport is optimized in the regime of adiabatically slow driving 
\cite{Bartussek}.
Very different from that, the anomalous subtransport is strongly suppressed  for slow
driving \cite{G10}. The physical mechanism behind this  nontrivial result is rooted in the asymptotic 
insensitivity of both the viscoelastic subdiffusion and subtransport to the presence of periodic potential
as it was discovered in Refs. \cite{Goychuk09,GH11,Goychuk12a}. The physical reason for this
astonishing feature can easily be understood by splitting the thermal fractional Gaussian  noise into 
two parts, fast
and slow, doing this self-consistently with respect to a characteristic time to escape out of 
the potential 
well. The latter time always
exists for the considered type of the anomalous stochastic dynamics. Fast noise components 
lead to
the formation of a non-Markovian rate (they kick out the particle out of potential well), while
the slow ones slowly modulate  this rate in time. Namely, this slow modulation leads to the emergence
of a large scale anti-persistency in the motion and determines  anomalous character of the
asymptotic dynamics.
For  very high potential barriers, the escape kinetics tends to the normal exponential one,
even though it remains
highly dispersive for the barriers smaller than about 10 $k_BT$, depending on $\alpha$ \cite{Goychuk09}. 
Surprisingly, the escape events 
are not transport limiting for such anomalously slow viscoelastic transport.
Subdiffusion cannot become faster in the presence of unbiased potential
than in its absence, and the asymptotic limit of potential-free subdiffusion is attained.
This feature of viscoelastic subdiffusion is very different from
the case of CTRW subdiffusion and fractional Fokker-Planck dynamics \cite{Metzler} in washboard potentials
that has been studied in \cite{GH06,H06}.
The given heuristic explanation is not only validated by stochastic numerical simulations in Refs.  
\cite{Goychuk09,GH11,Goychuk12a}, but is
also supported by two different analytical theories of quantum dissipative dynamics coupled to sub-Ohmic thermal
bath \cite{Chen,WeissBook}. The latter one presents a quantum generalization and counterpart 
of the classical FLE dynamics. 
These different quantum theories were
worked out only for a cosinusoidal potential $U(x)$ and only within a linear response regime
of transport. The results in Refs. \cite{Goychuk09,GH11,Goychuk12a} imply, however, that this 
remarkable feature (i)  
is purely classical,
i.e. no quantum-mechanical effects such as tunneling are involved, (ii) 
valid beyond the linear response regime,
(iii) is not restricted by cosinusoidal potentials but is valid for arbitrary periodic
potentials including ratchet potentials. In other terms, it is a truly universal feature, or
 universality
class of viscoelastic subdiffusion.

This universal property is responsible for vanishing rectification ratchet effect in the limit 
${\cal T}\to\infty$. However, for a constant driving force  the transition to this asymptotic limit is
generally  very slow and depends strongly on the amplitude of the periodic potential. 
This makes
the anomalous subdiffusive ratchets possible. Moreover, there exists an optimal driving frequency
since the rectification effect vanishes also in the limit of a very fast driving. It has been shown
recently \cite{GKh12a} for the ratchet potential
\begin{eqnarray}
U(x)=-U_0[\sin(2\pi x/L)+0.25 \sin(4\pi x/L) ] 
\end{eqnarray}
and harmonic driving, $f_{\rm ext}(t)=A\cos(\Omega t)$, that this optimization can be related to a stochastic
resonance (SR) effect \cite{Gammaitoni}, i.e. synchronization of the thermally activated 
jumps over one or several 
potential wells with
the half-period of the potential tilt in the transport direction. Such a \textit{non-Markovian} SR 
\cite{GH03,GH04,GH05} is clearly
impossible for anomalous dynamics based on divergent mean residence times. Moreover, the considered
dynamics fully retains inertial effects which can lead to weakly damped coherent oscillations within one
potential well, so that the conventional resonance is, in principle, also possible.

The role of the inertial effects can be characterized by
the dimensionless parameter $r_v=1/(\omega_0\tau_v)$, 
where $\omega_0=(2\pi/L)(3^{3/8}/2^{1/4}) \sqrt{U_0/m} $ is the bottom 
and (imaginary) top oscillation frequency in the considered potential at $A=0$, $f_0=0$, within the harmonic
approximation and in the absence of friction. 
The inertial effects can only be negligible 
for $r_v\ll 1$ and not too small $\alpha$ \cite{Burov}. 
The borderline value of $U_0$ corresponding to $r_v = 1$ is
in the dimensionless units  $U_0^*\approx 0.0157$. For all $U_0$ larger than this value, like
in this work, the inertial effects are very
essential. The corresponding frequency of damped oscillations $\omega_d$ can be found as the imaginary part
of the complex conjugated roots of the equation 
$s^2+\gamma_{\alpha}s^{\alpha}+\omega_0^2=0$,
where $\gamma_{\alpha}=\eta_{\alpha}/m$. From this in the used time scaling we obtain: 
$\omega_d\approx 4.17 $ for $U_0=0.25$, 
$\omega_d\approx 5.79$ for $U_0=0.50$, and $\omega_d\approx 7.04$ for $U_0=0.75$. For
the driving frequencies much smaller than $\omega_d$, as in this work, the conventional resonance 
effects are excluded, see but in Ref. \cite{KG12} for the case of flashing anomalous
ratchets, where they do play a profound role.

\begin{figure}[htbp]
\centerline{\includegraphics[scale=0.5]{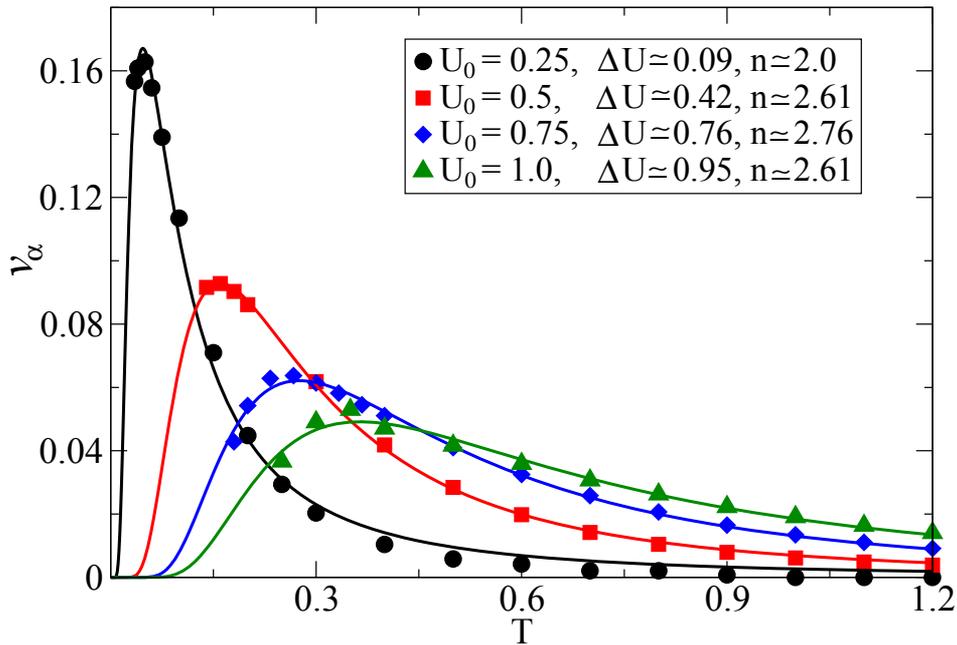}}
\caption{Subvelocity $v_\alpha$ \textit{vs.} temperature for $A=0.8$ and $\Omega=0.1$ 
and different potential
amplitudes $U_0$.}
\label{fig3}
\end{figure}

Previously, we have shown that anomalous ratchet transport optimizes at certain driving frequencies for a fixed driving 
amplitude, potential amplitude, and temperature \cite{G10}. It also optimizes versus 
temperature, for a fixed potential
amplitude, driving amplitude, and frequency \cite{GKh12a}, as also Fig. \ref{fig3} illustrates. 
Subvelocity shows a typical SR
dependence $v_\alpha(T)\propto T^{-n}\exp(-\Delta U/T)$ \cite{Gammaitoni}
with maximum at  $T_{\rm max}=\Delta U/n$. $\Delta U$
and $n$ are considered here as fitting parameters which deviate from the lowest second order response 
values $\Delta U=2 U_0$ and $n=2$. 

It should be also mentioned that  subdiffusion remains asymptotically  weakly sensitive to periodic driving.
The subdiffusion coefficient $D_\alpha$ does not significantly deviate from its potential and driving
free value $D_\alpha=T$ \cite{G10,GKh12a}. This means that for low temperatures and 
small $U_0$ the transport can be
low-dispersive and characterized by a large value of the generalized Peclet number  
${\rm Pe}_{\alpha}:=v_{\alpha}L/D_{\alpha}$. The latter quantity measures 
the coherence quality of the anomalous transport. 
Such a remarkable degree of coherence is also very different from the alternative
description of subdiffusion which is highly dispersive and featured by 
zero value of ${\rm Pe}_{\alpha}$ \cite{Goychuk12a}.

\begin{figure}[htbp]
\centerline{\includegraphics[scale=0.5]{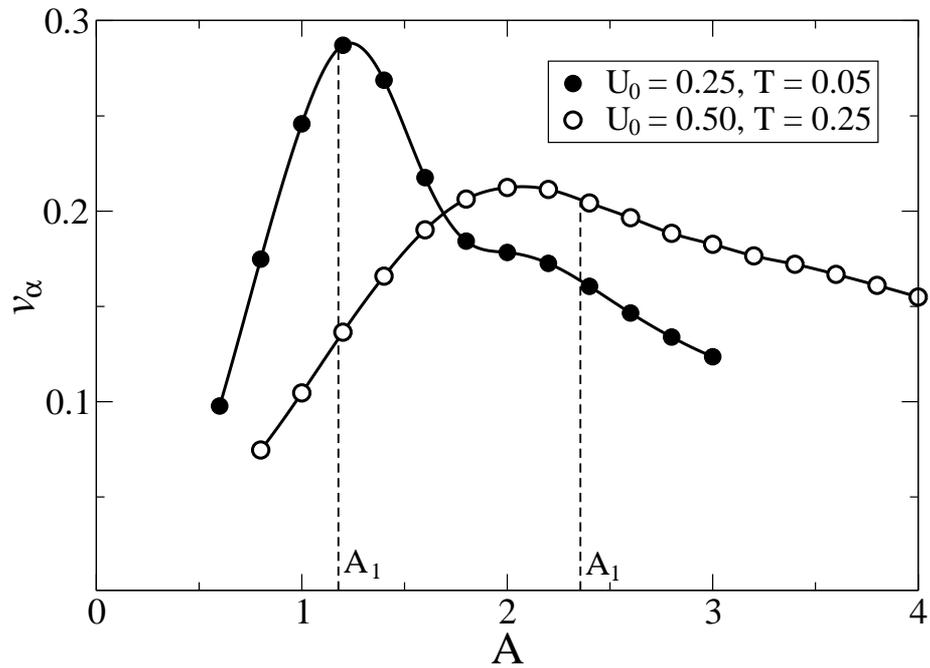}}
\caption{Subvelocity $v_\alpha$ as a function of the driving force amplitude $A$ for
 $\Omega=0.1$ and different values 
of the potential height $U_0$ and temperature $T$.}
\label{fig4}
\end{figure}

Furthermore, optimization occurs also with respect to driving strength. Indeed, for
small $A$, $\nu_\alpha \propto A^2$.
However, for $A\gg A_1=3 \pi U_0/(2L)$, where $A_1$ is the critical 
value of a static forward tilt at which the potential
barrier in the forward direction disappears, the transport should be suppressed. 
Therefore, one expects that it will
be optimal for the values of $A$ around $A_1$. 
Indeed, numerics in Fig. \ref{fig4} are consistent with this
prediction. 

\begin{figure}[htbp]
\centerline{\includegraphics[scale=0.5]{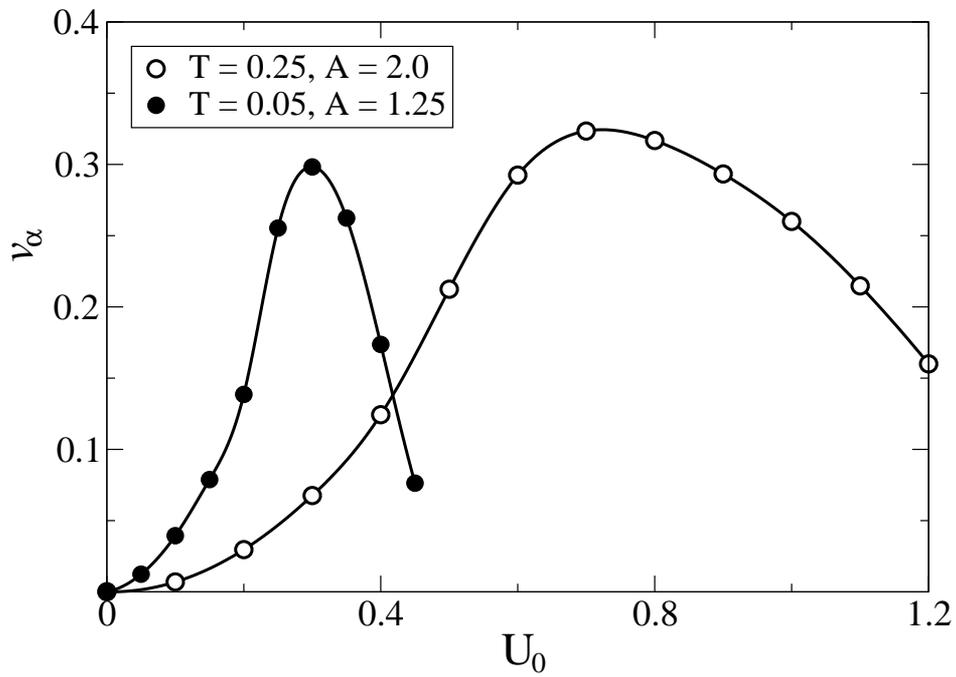}}
\caption{Subvelocity $v_\alpha$ as a function of the potential height $U_0$ at $\Omega=0.1$ and different values 
for driving amplitude $A$ and temperature $T$.}
\label{fig5}
\end{figure}

The transport should also be optimized with the potential amplitude $U_0$ for other parameters kept fixed.
The numerics in Fig. \ref{fig5} reveal this clearly too. One can conclude that anomalous 
ratchet transport can indeed be optimized with respect to any of parameters
$U_0$, $T$, $A$, $\Omega$. 

\section{Thermodynamic efficiency of anomalous ratchets}

The great interest presents also the dependence of the transport subvelocity $v_\alpha$
on the load $f_0$ in the direction opposite to transport 
and the thermodynamic efficiency of such a transport.
The first question has been answered recently in Ref. \cite{GKh12a}. 
For a broad range of parameters the dependence
of subvelocity on load turns out to be very simple,
\begin{equation}\label{load}
v_\alpha(f_0)\approx v_\alpha(0)-f_0,
\end{equation}
in the scaled dimensionless units. It must be stressed that the existence of a stopping force $f_{\rm cr}$
such that $v_\alpha(f_{\rm cr})=0$ makes it clear that we are dealing with a genuine ratchet 
effect to distinguish it from pseudo-ratchets, e.g. dissipationless  quantum ratchets, which
are featured by an infinitely small stopping  force and therefore cannot do any useful work. 
We shall answer below the second question, which was not addressed thus far for the 
anomalous ratchet transport. Here we 
follow
to the approach by Sekimoto \cite{Sekimoto}, and others \cite{Julicher} developed for the 
normal diffusion ratchets.
Our way of Markovian embedding makes its application to anomalous ratchets straightforward.

\begin{figure}[htbp]
\centerline{\includegraphics[scale=0.5]{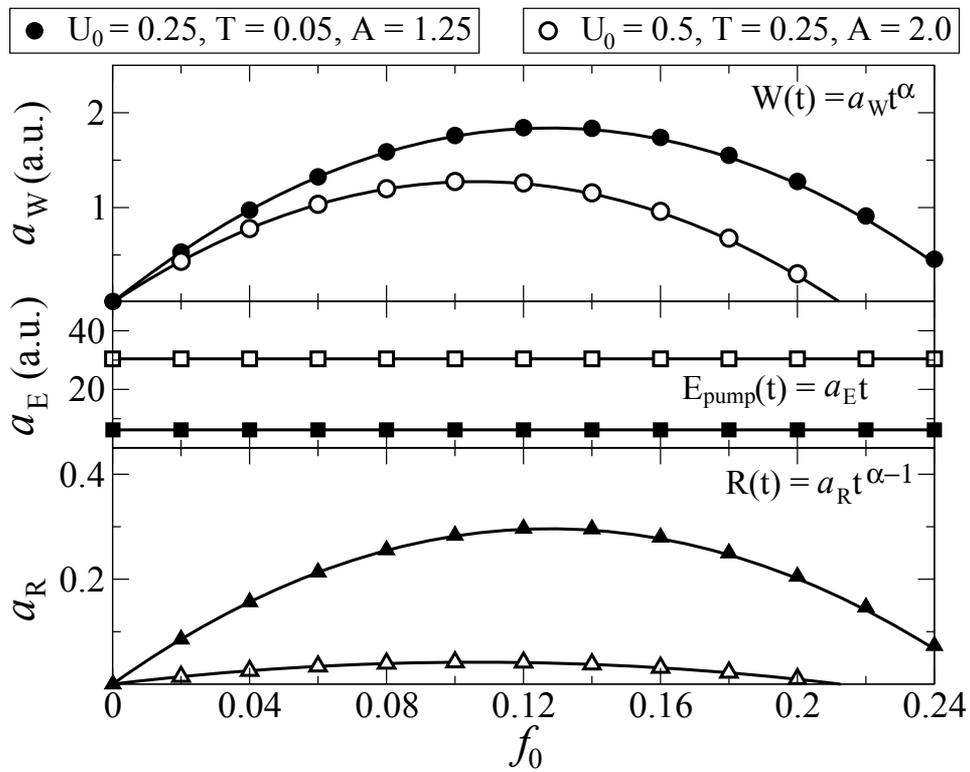}}
\caption{Dependencies of the coefficients for the useful work $a_W$, input energy $a_E$ and the 
efficiency $a_R$ versus load for two different sets of parameters and  a fixed value $\Omega=0.1$.
The symbols depict the numerical results and the solid lines show the corresponding analytical predictions.}
\label{fig6}
\end{figure}

Indeed, let us multiply the second equation in (\ref{embedding2}) by $v(t)$, integrate and
average it over the trajectory realizations.
The following energy balance equation follows:
\begin{eqnarray}
\Delta E_{\rm int}(t)=E_{\rm pump}(t)-W(t)+\Delta Q(t),
\end{eqnarray}
where $\Delta E_{\rm int}(t)=E_{\rm int}(t)-E_{\rm int}(0)$ is the change of internal energy of the Brownian 
motor particle, $E_{\rm int}(t)=\langle mv^2(t)/2+U(x(t))\rangle$,  within the time interval $[0,t)$, 
$E_{\rm pump}(t)=\int_0^t \langle f_{\rm ext}(t')v(t')\rangle dt'$ is the energy pumped by the
external force $f_{\rm ext}(t)$, $W(t)=f_0 \langle x(t)-x(0)\rangle$ is the useful work done against the 
load $f_0$,
and $\Delta Q(t)=\sum_{i=1}^N\int_0^t \langle u_i(t')v(t')\rangle dt'$ is the heat exchanged with the environment.
Obviously, $\Delta E_{\rm int}(t)$ is bounded and it can be neglected in the energy balance for large 
$t$. The efficiency of such an isothermal motor can be defined as 
$R(t)=W(t)/E_{\rm pump}(t)$. In principle (but not for the system considered!),
it can arrive at the theoretical maximum of one if isothermal motor operates near to the
thermal equilibrium, where the heat exchange with the thermal reservoir is absent, $\Delta Q\to 0$.
Then, however, also the operating power vanishes,
$P_W(t)=W/t\to 0$, i.e. the motor operates at zero power because of its operational speed is also
close to zero. Recent theoretical 
work shows that the maximal efficiency of isothermal engines at  maximal power can 
even exceed $50\%$ \cite{Broeck} as suggested in \cite{Seifert}.
The efficiency
of normal diffusion rocking ratchets is much lower, a few per cents only, at best. 
Subdiffusion introduces
new features. Namely, since the useful work done against the load scales linearly with the traveling
distance, it scales sublinearly with time, $W(t)=a_W t^\alpha$, 
where $a_W=f_0 v_\alpha(f_0)/\Gamma(1+\alpha)$, within the regime of subdiffusion. 
Therefore, the working power $P_W$ decays algebraically in time, $P_W\propto 1/t^{1-\alpha}$.
Like the concept of normal average velocity is modified for subvelocity in the case of
subtransport, one should modify  also the concept of power accordingly.  Let us do this by defining sub-power, 
$P_{\alpha,W}:=\Gamma(1+\alpha) W(t)/t^{\alpha}$, instead of the work produced
per unit of time.
In our notation, $P_{\alpha,W}=\Gamma(1+\alpha)a_W=f_0 v_\alpha(f_0)$.
However, the energy pumped into the system by the driving force scales linearly with
time, $\Delta E_{\rm pump}(t)=a_E t$, as numerical results  show.
This is also an expected result. Moreover, the pumping power $a_E$ does not depend on load, see in Fig. \ref{fig6}.
This is similar to the normal diffusion case \cite{Sekimoto}.
Therefore, the efficiency $R(t)$ declines
algebraically in time as $R(t)=a_R/t^{1-\alpha}$,  and for the coefficient $a_R$ we obtain,
upon taking into account Eq. (\ref{load}), a very simple result:
\begin{eqnarray}
a_R\approx f_0[v_\alpha(0)-f_0]/[a_E \Gamma(1+\alpha)] \;.
\end{eqnarray}
The maximal $a_R$ is achieved at $f_{\rm opt}=v_\alpha(0)/2$, i.e.
at the half of the stopping force $f_{\rm max}=v_\alpha(0)$.
Clearly, the efficiency is zero both for zero load and the
maximal load, and the maximal $a_R$ is
$a_R^{\rm (max)}=v_\alpha^2(0)/[4 a_E \Gamma(1+\alpha)]$.
This simple theory agrees with the numerical results in Fig. \ref{fig6} very well.
The efficiency decays algebraically slow and on appreciably long time scale it can 
compare well with the efficiency of normal diffusion ratchets. Indeed, the maximal
efficiency in Fig. \ref{fig6} for $t=1$ is about 30 \%, and for $t=100$ it is still 3 \%,
reducing to 0.03 \% at the end point $t_{\rm max}=10^6$ in our simulations. In this respect, the first
paper on the efficiency of normal diffusion rocking ratchets reported on the efficiencies as small
as 0.01 \% \cite{Sekimoto}. If $\alpha$ is closer to one, e.g. $\alpha=3/4$, as for diffusion in solutions of
semi-flexible polymers \cite{Amblard}, then the efficiency decays even slower, as $1/t^{1/4}$. 

The real molecular
motors do not operate as rocking ratchets, but rather as flashing ones with the minimum of the
potential which is displaced in the transport direction 
at each flash \cite{Julicher,Makhno}. The efficiency of such molecular motors can be
very high, close to maximally possible \cite{Julicher,Makhno}. How operating in such 
viscoelastic media as cytosol
of biological cells influences the efficiency of biological molecular motors is an intriguing issue
which we currently investigate. The results of this work point out into direction that
the efficiency of molecular motors can be also  high in the viscoelastic environments,
on the relevant mesoscopic  space/time scales.

\section{Summary and Conclusions}

In this work, we investigated anomalous ratchet effect within the Generalized Langevin Equation
approach to subdiffusion. Physically the origin of such a subdiffusion is rooted in the phenomenon of
viscoelasticity which leads to the emergence of long-range persistent anticorrelations in the 
Brownian particle displacements and also to negative correlations of its velocity fluctuations.
The power-law-decaying viscoelastic memory kernel corresponds to the sub-Ohmic model
of the thermal bath \cite{WeissBook} within the dynamical approach to generalized Brownian motion, 
where the
thermal environment is modeled by a quasi-infinite set of harmonic oscillators. 
In the inertialess limit, or on the time scale exceeding much the time scale of relaxation 
of the velocity variable and in the absence of external forces or for a constant force, 
this model yields fractional Brownian motion and therefore explains its dynamical origin.
Moreover, it is known  \cite{Goychuk07} that the dynamical response of such a motion trapped in parabolic
potentials is given, in neglecting the inertial effects, by the famous 
Cole-Cole expression \cite{Cole}. 
Such a response is indeed commonly measured in viscoelastic complex liquids and in 
glass-like media. It corresponds to the  $\beta-$relaxation. 
This provides an additional physical justification for the model considered. 
The corresponding memory
kernel can also be efficiently approximated by a sum of exponentials reflecting a hierarchy of
viscoelastic modes of environment. We used this fact for a highly efficient Markovian embedding
of the considered non-Markovian  GLE dynamics with long-range memory. This methodology allows also for
a vivid physical interpretation. Namely, the medium is represented by a set of inertialess Brownian
quasi-particles elastically coupled to the moving Brownian particle (see video on the web site
of journal). Fast medium's deformation
(represented by rapidly relaxing quasi-particles) follows immediately to the Brownian particle,
which effectively becomes dressed
with these fast Brownian particles. This reminds the small polaron concept in 
condensed matter physics.
However, the slow deformations cannot immediately follow to the particle. Their retarded motion provides 
a physical mechanism for slowly
decaying memory about the former positions of the Brownian particle and causes subdiffusion on a long
time scale. The presence of a periodic potential influences the motion of Brownian particle, but it
does not affect directly the medium. As a result, the dynamics of slow deformations is not affected
by the presence of potential and this leads to the universality class of viscoelastic
subdiffusion and transport in tilted periodic potentials. Here, the asymptotic behavior of the Brownian 
particles does not depend on the presence of periodic potential. For this reason, the 
anomalous ratchet effect
is not possible for adiabatically slow rocking.
 However, it emerges for a sufficiently fast driving. Moreover, it optimizes when the driving period
matches a characteristic time scale of the temperature 
activated transitions to the neighboring potential wells.
This leads to the optimization of anomalous ratchet transport with respect to different parameters:
frequency and amplitude of the time-periodic driving force, potential amplitude, and temperature.
The last three optimizations were addressed in this paper. We studied also the problem of thermodynamic
efficiency of such anomalous isothermal Brownian motors and proposed a simple theory which remarkably
well agrees with the numerical results. Most importantly, since the position of Brownian particle
scales sublinearly with time, one cannot characterize such anomalous Brownian motors using the
conventional notion of working power, since the useful work done by motors against a load scales also
sublinearly with time. Instead, we introduced the notion of sub-power and showed that the efficiency
of such motors decays algebraically slow in time since the energy pumped into the directed motion scales with
time linearly. Together with our other recently published works this contribution opens the new field
of anomalous Brownian motors which can especially be relevant for transport processes in living
cells, in a more realistic setup of flashing ratchets pertinent to biological 
molecular motors \cite{Julicher}. This future direction grows naturally
from the research already done. It will be explored in the nearest future.

\vspace*{0.5cm}

\begin{acknowledgement}
Support of this research by the Deutsche Forschungsgemeinschaft, Grants
GO 2052/1-1 and GO 2052/1-2 is gratefully acknowledged.
\end{acknowledgement}



\end{document}